\documentclass{article}
\usepackage[preprint]{spconf}
\usepackage{amsmath,graphicx}
\usepackage{color}
\usepackage{subcaption}
\usepackage{multicol}
\usepackage{amsmath,bm}
\usepackage{booktabs}
\usepackage{paralist}
\usepackage{hyperref}
\usepackage{cite}

\copyrightnotice{\copyright\ IEEE 2018}

\def\vec#1{\ensuremath{\boldsymbol{{#1}}}}

\newcommand{\TS}[1]{\textcolor{red}{\bf \small [ #1 --TS]}}
\newcommand{\RW}[1]{\textcolor{blue}{\bf \small [ #1 --RW]}}
\newcommand{\BL}[1]{\textcolor{cyan}{\bf \small [ #1 --BL]}}
\newcommand{\ST}[1]{\textcolor{magenta}{\bf \small [ #1 --ST]}}
\renewcommand{\TS}[1]{}
\renewcommand{\RW}[1]{}
\renewcommand{\BL}[1]{}
\renewcommand{\ST}[1]{}

\title{Multilingual Speech Recognition with a Single End-to-End Model}
\twoauthors{Shubham Toshniwal\sthanks{Work done at Google NYC.}}
        {Toyota Technological Institute at Chicago\\ \small{\texttt{shtoshni@ttic.edu}}}
        {Tara N. Sainath, Ron J. Weiss, Bo Li,}
        {Pedro Moreno, Eugene Weinstein, Kanishka Rao}
        {Google Inc., U.S.A\\  \small{\texttt{ \{tsainath, ronw, boboli, }}}
        {\small{\texttt{pedro, weinstein, kanishkarao\}@google.com}}}

\begin{document}
\ninept
\maketitle
\begin{abstract}
  Training a conventional automatic speech recognition (ASR) system to support multiple languages is challenging because the sub-word unit, lexicon and word inventories are typically language specific.
  In contrast, sequence-to-sequence models are well suited for multilingual ASR because they encapsulate an acoustic, pronunciation and language model jointly in a single network.
  In this work we present a single sequence-to-sequence ASR model trained on 9 different Indian languages, which have very little overlap in their scripts.
  Specifically, we take a union of language-specific grapheme sets and train a grapheme-based sequence-to-sequence model jointly on data from all languages.
  We find that this model, which is not explicitly given any information about language identity, improves recognition performance by 21\% relative compared to analogous sequence-to-sequence models trained on each language individually.
  By modifying the model to accept a language identifier as an additional input feature, we further improve performance by an additional 7\% relative and eliminate
  confusion between different languages.
\end{abstract}
\begin{keywords}
    ASR, speech recognition, multilingual, encoder-decoder, seq2seq, Indian
\end{keywords}

\section{Introduction}
\label{sec:intro}

Speech recognition has made remarkable progress in the past few years with services such as Google Voice Search supporting about 120 languages.\footnote{\url{https://www.blog.google/products/search/type-less-talk-more/}}
Further expanding its coverage of the world's $\approx\,$7,000 languages is of great interest to both academia and industry. However, in many cases the resources available to train large vocabulary continuous speech recognizers are severely limited~\cite{Besacier14}.
These challenges have meant that there has been a perennial interest in multilingual and cross-lingual models which allow for knowledge transfer across languages, and thus relieve burdensome data requirements~\cite{Weng97,Schultz97, Schultz01,Niesler07,Lin09,burget10, Thomas12, Heigold13, Ghoshal13, huang13, Watanabe17}.

Most of the previous work on multilingual speech recognition has been limited to making the acoustic model (AM) multilingual~\cite{Schultz97, Schultz01, Niesler07, Lin09, Heigold13, Ghoshal13, huang13, Vu14, Tong17}.
Some of the multilingual AMs require a common phone set~\cite{Schultz97, Schultz01, Vu14} while others share some of the acoustic model parameters~\cite{Heigold13, Ghoshal13, huang13, Chen15}.
A hat swap structure is proposed in \cite{Heigold13, Ghoshal13, huang13}, where the lower layers of a deep neural network (DNN) are shared across languages and the output layer is language-specific.
Alternatively, multilingual bottleneck features from a DNN feature extractor can be used for either a Gaussian Mixture Model or DNN-based systems \cite{tuske2013investigation}.
These multilingual AMs still require language-specific pronunciation models (PMs) and language models (LMs) which means that often such models must know the speech language identity during inference~\cite{Heigold13, Ghoshal13, huang13}.
Moreover, the AMs, PMs and LMs are usually optimized independently, in which case errors from one component propagate to subsequent components in a way that was not seen during training.

Sequence-to-sequence models fold the AM, PM and LM into a single network, making them attractive to explore for multilingual speech recognition. Building a multilingual sequence-to-sequence model requires taking the union over all the language-specific grapheme sets and training the model jointly on data from all the languages.
In addition to their simplicity, the end-to-end nature of such models means that all of the model parameters contribute to handling the variations between different languages.
Our attention-based sequence-to-sequence model is based on the Listen, Attend and Spell (LAS) model~\cite{Chan16, Prabhavalkar17}, the details of which are explained in the next section.
%
Our work is most similar to that of \cite{Watanabe17} which similarly
proposes an end-to-end trained multilingual recognizer to directly
predict grapheme sequences in 10 distantly related languages.
They utilize a
hybrid attention/connectionist temporal classification model
integrated with an independently trained grapheme LM.
In this paper we use a simpler sequence-to-sequence model without
an explicit LM, and study a corpus of 9 more closely related Indian
languages.

We show that a LAS model jointly trained across data from 9 Indian languages without any explicit language specification consistently outperforms monolingual LAS models trained independently on each language.
Even without explicit language specification, the model is rarely confused between languages.
We also experiment with certain language-dependent variants of the model.
In particular, we obtain the largest improvement by conditioning the encoder on the speech language identity.
We also run several experiments on synthesized data to gain insights into the behavior of these models.  We find that the multilingual model is unable to code-switch between languages, indicating that the language model is dominating the acoustic model.
Finally, we find that the language-conditioned model is able to transliterate Urdu speech into Hindi text, suggesting that the model has learned an internal representation which disentangles the underlying acoustic-phonetic content from the language.

\section{Model}
\vspace{-0.07in}
In this section we describe the Listen, Attend and Spell (LAS) attention-based sequence-to-sequence ASR model proposed by Chan et al~\cite{Chan16}, as well as our proposed modifications to support recognition in multiple languages.

\subsection{LAS Model}
The sequence-to-sequence model consists of three modules: an \emph{encoder}, \emph{decoder} and \emph{attention network} which are trained jointly to predict a sequence of graphemes from a sequence of acoustic feature frames.

We use 80-dimensional log-mel acoustic features computed every 10ms over a 25ms window. Following \cite{Sak15} we stack 8 consecutive frames and stride the stacked frames by a factor of 3.
This downsampling enables us to use a simpler encoder architecture than \cite{Chan16}.

The encoder is comprised of a stacked bidirectional recurrent neural network (RNN)~\cite{Hochreiter97, Schuster97} that reads acoustic features $\vec{x} = (\vec{x}_1, \dots, \vec{x}_K$) and
outputs a sequence of high-level features (hidden states) $\vec{h}$ = ($\vec{h}_1, \dots, \vec{h}_K$).
The encoder is similar to the acoustic model in an ASR system.

The decoder is a stacked unidirectional RNN that computes the probability of a sequence of
characters $\vec{y}$ as follows:
\begin{equation*}
P(\vec{y}|\vec{x}) = P(\vec{y}|\vec{h}) = \prod_{t=1}^{T}P(y_{t}|\vec{h}, \vec{y_{< t}}).
\end{equation*}

The conditional dependence on the encoder state vectors $\vec{h}$ is represented
 by context vector $\vec{c}_{t}$, which is a function of the current decoder
hidden state and the encoder state sequence:
\begin{align*}
    u_{it} &= \vec{v}^\top \tanh(\vec{W_h}\vec{h}_i + \vec{W_d}\vec{d}_t + \vec{b_\text{a}}) \\
\vec{\alpha}_{t} & = \text{softmax}(\vec{u}_t) \qquad \vec{c}_t = \sum_{i=1}^{K} \alpha_{it}\vec{h}_{i}
\end{align*}
where the vectors $\vec{v}, \vec{b_\text{a}}$ and the matrices $\vec{W_h}, \vec{W_d}$
are learnable parameters; $\vec{d}_t$ is the hidden state of the decoder at time
 step $t$.

The hidden state of the decoder, $\vec{d}_{t}$, which captures the previous
character context $\vec{y_{< t}}$, is given by:
$$\vec{d}_{t} = \text{RNN}(\tilde{\vec{y}}_{t-1}, \vec{d}_{t-1}, \vec{c}_{t-1})$$
where $\vec{d}_{t-1}$ is the previous hidden state of the decoder, and $\vec{\tilde{y}}_{t-1}$ is a
character embedding vector for $y_{t-1}$, as is typical practice in RNN-based
language models.
The decoder is analogous to the language model component of a pipeline system for ASR.
The posterior distribution of the output at time step $t$ is given by:
$$P(y_{t}|\vec{h}, \vec{y_{< t}}) = \text{softmax}(\vec{W_\text{s}}[\vec{c}_{t}; \vec{d}_{t}] + \vec{b_\text{s}}),$$
where $\vec{W_\text{s}}$ and $\vec{b_\text{s}}$ are again learnable parameters.
%
The model is trained to optimize the discriminative loss:
$$L_\text{LAS} = -\log(P(\vec{y}|\vec{x}))$$

\subsection{Multilingual Models}
In the multilingual scenario, we are given $n$ languages $\{\mathcal{L}_1, \dots, \mathcal{L}_n\}$, each with independent character sets $\{\mathcal{C}_1, \mathcal{C}_2, \cdots, \mathcal{C}_n\}$ and
training sets $\{(\mathcal{X}_1, \mathcal{Y}_1), \dots, (\mathcal{X}_n, \mathcal{Y}_n)\}$.
The combined training dataset is thus given by the union of the datasets for each language:
$$(\mathcal{X}, \mathcal{Y}) = \cup_{i=1}^n (\mathcal{X}_i, \mathcal{Y}_i)$$
and the character set for the combined dataset is similarly given by:
$$\mathcal{C} = \cup_{i=1}^{n}\mathcal{C}_i$$

\subsubsection{Joint}
We begin by training a joint model, consisting of the LAS model described in the previous section trained directly on the combined multilingual dataset.
This model is not given any explicit indication that the training dataset is composed of different languages.
However, as we will show later, this model is still able to recognize speech in multiple languages despite the lack of runtime language-specification.

\subsubsection{Multitask}
We also experiment with a variant of the joint model which has the same architecture but is trained in a multitask learning (MTL) configuration~\cite{Caruana97} to jointly recognize speech and simultaneously predict its language.
The language ID annotation is thus utilized during training, but is not passed as an input during inference.
In order to predict the language ID, we average the encoder output $h$ across all time frames 
to compute an utterance-level feature.
This averaged feature is then passed to a softmax layer to predict the likelihood of the speech belonging to each language:
\begin{align*}
      p(\mathcal{L} | \vec{x}) &= \text{softmax}(\vec{W_\text{lang}} \, \tfrac{1}{K}\Sigma_i \vec{h}_i + \vec{b_\text{lang}})
\end{align*}
The language identification loss is given by:
$$L_\text{LID} = - \log(p(\mathcal{L}=\mathcal{L}_j | \vec{x})$$
where the $j$-th language, $\mathcal{L}_j$, is the ground truth language.
The two losses are combined using an empirically determined weight $\lambda$ to obtain the final training loss:
$$L_\text{MTL} = \tfrac{1}{1 + \lambda} L_\text{LAS} + \tfrac{\lambda}{1 + \lambda} L_\text{LID}$$

\subsubsection{Conditioned}
Finally, we consider a set of conditional models which utilize the language ID during inference.
Intuitively, we expect that a model which is explicitly conditioned on the speech language will  have an easier time allocating its capacity appropriately across languages, speeding up training and improving recognition performance.

Specifically,  we learn a fixed-dimensional language embedding for each language to condition different components of the basic joint model on language ID.
This conditioning is achieved by feeding in the language embedding as an input to the first layer of encoder, decoder or both giving rise to
\begin{inparaenum}[(a)]
\item \textit{Encoder-conditioned},
\item \textit{Decoder-conditioned}, and
\item \textit{Encoder+Decoder-conditioned}
\end{inparaenum}
variants.
In contrast to the MTL model, the language ID is not used as part of the training cost.

\section{Experimental Setup}

\begin{table}[h!]
\centering
\caption{Multilingual dataset statistics.}
\label{data_split}
    \begin{tabular}{lll}
    \toprule
    Language & \# training utts. & \# test utts. \\
    \midrule
    Bengali 
             & 364617            & 14679         \\
    Gujarati 
             & 243390            & 14935         \\
    Hindi 
             & 213753            & 14718         \\
    Kannada 
             & 192523            & 14765         \\
    Malayalam 
             & 285051            & 14095         \\
    Marathi 
             & 227092            & 13898         \\
    Tamil 
             & 164088            & 9850          \\
    Telugu 
             & 232861            & 14130         \\
    Urdu 
             & 196554            & 14486         \\
    \midrule
    Total & 2119929 & 125556 \\
    \bottomrule
\end{tabular}
\label{tab:data}
\end{table}

\subsection{Data}
We conduct our experiments on data from nine Indian languages shown in Table~\ref{tab:data}, which corresponds to a total of about 1500 hours of training data and 90 hours of test data.
The nine languages have little overlap in their character sets, with the exception of Hindi and Marathi which both use the Devanagari script.
The small overlap means that the output vocabulary for our multilingual models, which is union over character sets, is also quite large, containing 964 characters.
Separate validation sets of around 10k utterances per language are used for hyperparameter tuning.
All the utterances are dictated queries collected using desktop and mobile devices.
\subsection{Model and Training Details}
As a baseline, we train nine monolingual models independently on data for each language.  We tune the hyperparameters on Marathi and reuse the optimal configuration to train models for the remaining languages.
The best configuration for Marathi uses a 4 layer encoder comprised of 350 bidirectional long short-term memory (biLSTM) cells (i.e.\ 350 cells in forward layer and 350 cells in backward layer), and
a 2 layer decoder containing 768 LSTM cells in each layer.
For regularization, we apply a small L2 weight penalty of 1e-6 and add Gaussian weight noise ~\cite{Graves11} with standard deviation of 0.01 to all parameters after 20k training steps.
All the monolingual models converge within 200-300k gradient steps.

Since the multilingual training corpus is much larger, we were able to train a joint larger multilingual model without overfitting.
As with the training set, the validation set is also a union of the language-specific validation sets.
The best configuration uses a 5 layer encoder comprised of 700 biLSTM cells, and
a 2 layer decoder containing 1024 LSTM cells in each layer.
For the multitask model, we find $\lambda=0.01$ among $\{0.1, 0.01\}$ to work the best. 
We restricted ourselves to these values because for a very large $\lambda$, the language ID prediction task would dominate the primary task of ASR, while for a very small $\lambda$ the additional task would have no effect on the training loss.
For all conditional models, we use a 5-dimensional language embedding.
For regularization we add Gaussian weight noise with standard deviation of 0.0075 after 25k training steps.
All multilingual models are trained for approximately 2 million steps.

All models are implemented in TensorFlow~\cite{tensorflow} and
trained using asynchronous stochastic gradient descent~\cite{Dean12} using 16 workers. 
The initial learning rate is set to 1e-3 for the monolingual models and 1e-4 for the multilingual models with learning rate decay in all the models.

\section{Results}

\begin{table}[h]
\centering
\caption{WER(\%) of language-specific, joint, and joint+MTL LAS models.}
\begin{tabular}{lccc}
        \toprule
        Language      & Language-specific          & Joint    & Joint + MTL           \\
        \midrule
        Bengali 
                      & 19.1	            & 16.8              & {\bf 16.5}    \\
        Gujarati 
                      & 26.0                & {\bf 18.0}        & 18.2          \\
        Hindi 
                      & 16.5                & {\bf 14.4}        & {\bf 14.4}    \\
        Kannada 
                      & 35.4                & {\bf 34.5}        & 34.6          \\
	Malayalam 
                      & 44.0                & 36.9              & {\bf 36.7}    \\
        Marathi 
                      & 28.8                & 27.6              & {\bf 27.2}    \\
        Tamil 
                      & 13.3                & 10.7              & {\bf 10.6}    \\
        Telugu 
                      & 37.4                & {\bf 22.5}        & 22.7          \\
        Urdu 
                      & 29.5                & 26.8              & {\bf 26.7}    \\
        \midrule
        Weighted Avg. & 29.05               & {\bf 22.93}       & {\bf 22.91}   \\
        \bottomrule
\end{tabular}
\label{tab:mum}
\end{table}

\BL{maybe include the weight used for each language for the ``Weighted Avg.''}

We first compare the language-specific LAS models with the joint LAS model trained on all languages.
As shown in Table~\ref{tab:mum}, the joint LAS model outperforms the language-specific models for all the languages.
In fact, the joint model decreases weighted average WERs across all the 9 languages, weighted by number of words, by more than 21\% relative to the monolingual models.
This result is quite interesting not only because the joint model is a single model that is being compared to 9 different monolingual models, but unlike the monolingual models the joint model it not language-aware at runtime.
Finally, the large performance gain of the joint model is also attributable to the fact that
the Indian languages are very similar in the phonetic space~\cite{Lavanya05}, despite
using different grapheme sets.

Second, we compare the joint LAS model with the multitask trained variant.
As shown in the right two columns of Table~\ref{tab:mum}, the MTL model shows limited improvements over the joint model.
This might be due to the following reasons:
\begin{inparaenum}[(a)]
\item Static choice of $\lambda$.
    Since the language ID prediction task is easier than ASR, a dynamic $\lambda$ which is high initially and decays over time might be better suited, and
\item The language ID prediction mechanism of averaging over encoder outputs might not be ideal.
  A learned weighting of the encoder outputs, similar to the attention module, might be better suited for the task.
\end{inparaenum}

\begin{table}[h]
\centering
\caption{WER(\%) of joint LAS model and the joint language-conditioned models, namely decoder-conditioned (Dec), encoder-conditioned (Enc), and encoder+decoder-conditioned (Enc + Dec).}
\begin{tabular}{lcccc}
        \toprule
        Language  & Joint & Dec   & Enc   & Enc + Dec   \\
        \midrule
        Bengali 
                  & 16.8          & 16.9      & {\bf 16.5}        & {\bf 16.5}            \\
	Gujarati 
                  & 18.0          & 17.7      & {\bf 17.2}        & 17.3                  \\
        Hindi 
	          & {\bf 14.4}    & 14.6      & 14.5              & {\bf 14.4}            \\
        Kannada 
	          & 34.5          & 30.1      & 29.4              & {\bf 29.2}            \\
        Malayalam 
	          & 36.9          & 35.5      & 34.8              & {\bf 34.3}            \\
        Marathi 
	          & 27.6          & 24.0      & {\bf 22.8}        & 23.1                  \\
        Tamil 
	          & 10.7          & 10.4      & {\bf 10.3}        & 10.4                  \\
        Telugu 
	          & 22.5          & 22.5      & 21.9              & {\bf 21.5}            \\
        Urdu 
	          & 26.8          & 25.7      & {\bf 24.2}        & 24.5                  \\
        \midrule
        Weighted Avg.& 22.93        & 22.03     & 21.37             & {\bf 21.32}            \\
        \bottomrule
\end{tabular}
\label{tab:cond}
\end{table}

\begin{figure*}[t!]
\begin{subfigure}{.5\textwidth}
  \centering
    \includegraphics[width=.9\linewidth]{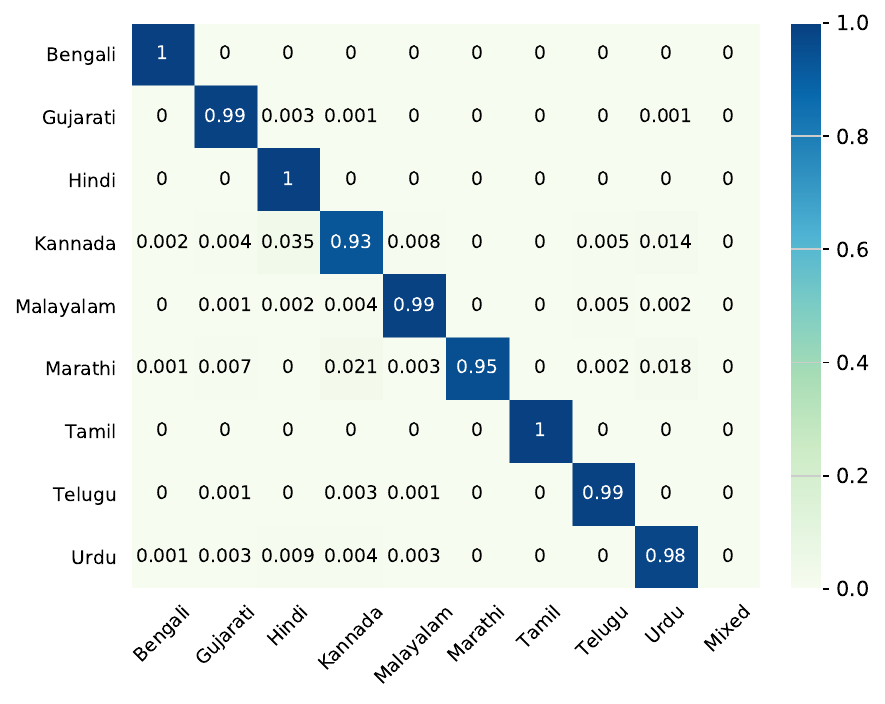} 
  \caption{Joint}
  \label{fig:uncon_conf}
\end{subfigure}%
\begin{subfigure}{.5\textwidth}
  \centering
    \includegraphics[width=.9\linewidth]{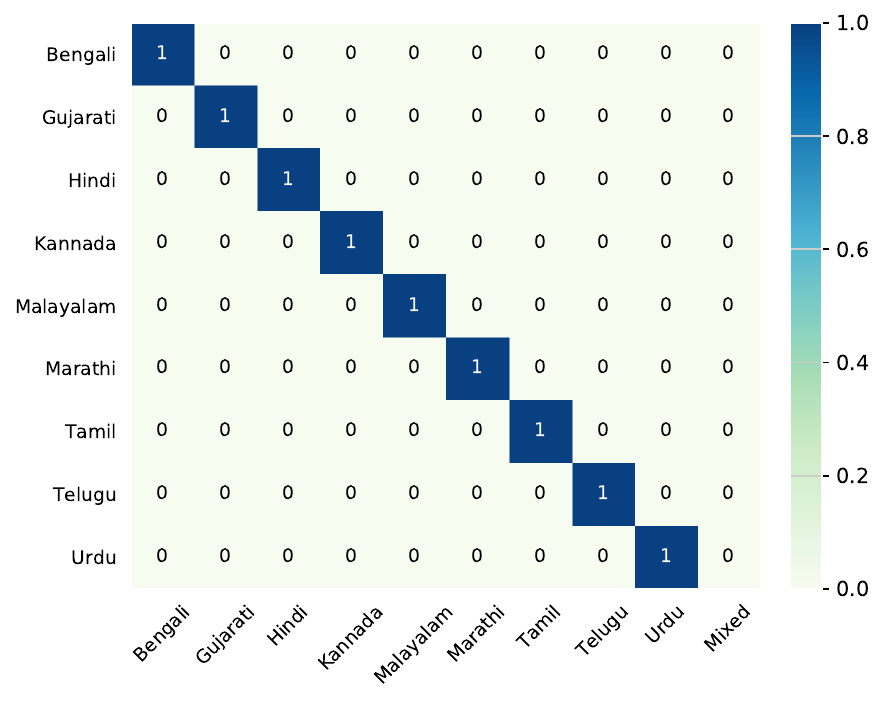} 
  \caption{Encoder-conditioned}
  \label{fig:con_conf}
\end{subfigure}
\caption{Confusion matrices for joint and encoder-conditioned models, truncated to precision of $10^{-3}$. The joint model is rarely confused between languages, while conditioning removes those rare cases almost completely.}
\label{fig:conf_mat}
\end{figure*}

Third, Table~\ref{tab:cond} shows that all the joint models conditioned on the language ID outperform the joint model.
The encoder-conditioned model (Enc) is better than the decoder-conditioned model (Dec) indicating that some form of acoustic model adaptation towards different languages and accents occurs when the encoder is conditioned.
In addition, conditioning both the encoder and decoder (Enc + Dec) does not improve much over conditioning just the encoder, suggesting that feeding the encoder with language ID information is sufficient, as the encoder outputs are then fed to the decoder anyways via the attention mechanism.

Comparing model performances across languages we see that all the models perform worst on Malayalam and Kannada.
We hypothesize that this has to do with the \emph{agglutinative} nature of these languages which makes the average word longer in these languages compared to languages like Hindi or Gujarati.
For example, an average training set word in Malayalam has 9 characters compared to 5 in Hindi.
In fact, we found that in contrast to the WER, the character error rate (CER) for Hindi and Malayalam were quite close.

\section {Analysis}
In this section we investigate the behavior and capacity of the proposed system in more detail, by asking the questions detailed below.\\[-0.5em]

{\bf How often does the model confuse between languages?} The ability of the proposed model to recognize multiple languages comes with the potential side effect of confusing the languages.
The lack of script overlap between Indian languages, with the exceptions of Hindi and Marathi, means that the surface analysis of the script used in the model output is a good proxy to tell if the model is confused between languages or not.
We carry out this analysis at the word level and
 check if the output words use graphemes from a single language or a mixture.
We test the word first on the ground truth language, and in case of failure, test it on other languages.
If the word cannot be expressed using the character set of any single language, we classify it as \emph{mixed}.
The result for both the joint and the encoder-conditioned model is summarized in Figure~\ref{fig:conf_mat}.
While both the models are rarely confused between languages, the result for the joint model is interesting given its lack of explicit language awareness, showing that the LAS model is implicitly learning to predict language ID. It is also interesting to observe that by conditioning the joint model on the language ID,
there is no confusion between languages.\\[-0.5em]

\noindent {\bf Can the joint model perform code-switching?} The joint model in theory has the capacity to switch between languages. In fact, it can code-switch between English and the 9 Indian languages due to the presence of English words in the training data\footnote{1-6\% of the total words in the training set are English words in all the 9 languages.}.
We were interested in testing if the model could also code-switch between a pair of Indian languages which was not seen during training.
For this purpose, we created an artificial dataset by selecting about 1,000 Tamil utterances and appending them with the same number of Hindi utterances with a 50ms break in between.
To our disappointment, the model is not able to code-switch at all.
It picks one of the two scripts and sticks with it.
Manual inspection shows that: (a) when the model chooses Hindi, it only transcribes the Hindi part of the utterance (b) similarly when the model chooses Tamil it only transcribes the Tamil part, but on rare occasions it also transliterates the Hindi part. This suggests that the language model is dominating the acoustic model and points to overfitting, which is a known issue with attention-based sequence-to-sequence models~\cite{Chorowski17}.\\[-0.5em]

\noindent {\bf What does the conditioned model output for mismatched language ID?}
The interesting question here is does the model obey acoustics or is it faithful to the language ID.
To answer this, we created an artificial dataset of about 1,000 Urdu utterances labeled with the Hindi language ID and transcribed it with the encoder-conditioned model.
As it turns out, the model is extremely faithful to the language ID and sticks to Hindi's character set.
Manual inspection of the outputs reveals that the model transliterates Urdu utterances in Hindi, suggesting that the model has learned an internal representation which disentangles the underlying acoustic-phonetic content from the language identity.

\vspace{-0.07in}
\section{conclusion}
\vspace{-0.07in}
We present a sequence-to-sequence model for multilingual speech recognition which is able to recognize speech without any explicit language specification. We also propose simple variants of the model conditioned on language identity. 
The proposed model and its variants substantially outperform baseline monolingual sequence-to-sequence models for all languages, and rarely chooses the incorrect grapheme set in its output.
The model, however, cannot handle code-switching, suggesting that the language model is dominating the acoustic model. 
In future work, we would like to integrate the conditional variants of the model with separate language-specific language models to further improve recognition accuracy. 
We would also like to compare the proposed models against traditional models on live traffic data. 
The exploration of reasons for lack of code-switching in joint model can also lead to interesting insights regarding sequence-to-sequence models.

\vspace{-0.07in}
\section{Acknowledgements}
\vspace{-0.07in}
We would like to thank Rohit Prabhavalkar, Yonghui Wu, Vijay Peddinti, Zhifeng Chen and Patrick Nguyen for helpful comments.
We are also thankful to the anonymous reviewers for their helpful comments.

\bibliographystyle{IEEEbib}
\bibliography{references}

\end{document}